\documentclass[10pt,prd,longbibliography,superscriptaddress]{revtex4-2}
\usepackage[T1]{fontenc}
\usepackage[english]{babel}
\usepackage{gensymb}

\usepackage{amsfonts,amsbsy,amssymb,amsmath,graphicx,float} 
\usepackage{color}

\graphicspath{{figures/}}

\begin{document}

\title{Bifurcation of dividing surfaces constructed from period-doubling bifurcations of periodic orbits in a caldera potential energy surface}

\author{Matthaios Katsanikas}
\email{mkatsan@academyofathens.gr}
\affiliation{Research Center for Astronomy and Applied Mathematics, Academy of Athens, Soranou Efesiou 4, Athens, GR-11527, Greece.}
\affiliation{School of Mathematics, University of Bristol, \\ Fry Building, Woodland Road, Bristol, BS8 1UG, United Kingdom.}

\author{Makrina Agaoglou}
\email{makrina.agaoglou@icmat.es}
\affiliation{Instituto de Ciencias Matem{\'a}ticas, CSIC, C/Nicol{\'a}s Cabrera 15, Campus Cantoblanco,
28049 Madrid, Spain}

\author{Stephen Wiggins}
\email{s.wiggins@bristol.ac.uk}
\affiliation{School of Mathematics, University of Bristol, \\ Fry Building, Woodland Road, Bristol, BS8 1UG, United Kingdom.}

\begin{abstract}
In this work we analyze the bifurcation of dividing surfaces that occurs as a result of  two period-doubling bifurcations in a 2D caldera-type potential. We  study the structure, the range, the minimum and maximum extent  of the  periodic orbit dividing surfaces  before and after a subcritical  period-doubling bifurcation of the family of the central minimum of the potential energy surface. Furthermore we  repeat the same study for the case for a supercritical   period-doubling bifurcation of the family of the central minimum of the potential energy surface. We will discuss and compare the results for  the two case of bifurcations of dividing surfaces.    
\end{abstract}

\maketitle


\section{Introduction}

The aim of this paper is to study a bifurcation
of periodic orbit dividing surfaces that occurs as a result of a period-doubling bifurcation of periodic orbits. We studied the structure, the range, and the minimum and maximum of the dividing surfaces. We study this kind of bifurcation in a caldera type system \cite{carpenter1985,collins2013nonstatistical}. The original form of this potential is characterized by one central minimum and four index-1 saddles around it that control the exit and entrance to four wells \cite{carpenter1985} or to the infinity \cite{collins2013nonstatistical}. This kind of systems is encountered in many organic chemical reactions and it has been studied  recently in many papers \cite{collins2013nonstatistical,katsanikas2020b,katsanikas2020a,katsanikas2018phase,katsanikas2019phase,geng2021a,geng2021influence}. Similar type of potentials can be encountered in four-armed barred galaxies \cite{athanassoula2009rings}.

The dividing surfaces  play important role in the transition state theory (TST)  \cite{Wigner38,waalkens2007} in chemical reaction dynamics \cite{Wigner38,waalkens2007}, in dynamical astronomy \cite{reiff2022stability} and in fluid dynamics \cite{Bottaro}. These surfaces are one less dimension than that of the potential energy surface. This means that in Hamiltonian systems with two degrees the dividing surfaces are 2-dimensional objects. We constructed these objects using the classical algorithm of \cite{Pechukas73, Pechukas77,Pechukas79,pollak1985,pechukas1981}. This method is valid only in Hamiltonian systems with two degrees of freedom. The construction of these objects can be done in Hamiltonian systems with three or more degrees of freedom, using a higher dimensional object, the  Normally Hyperbolic Invariant Manifold -NHIM (see for example \cite{wiggins2016} and references therein). The construction of these objects, using periodic orbits,  in Hamiltonian systems with three or more degrees of freedom has been done recently by \cite{katsanikas2021ds,katsanikas2021bds}. 

In the past, many researchers  studied the bifurcations of the transition states (periodic orbits and NHIMs) that are the basic elements for the construction of dividing surfaces  \cite{burghardt1995molecular,Li09,Inarrea11,founargiotakis1997bifurcation,katsanikas2020c,Agaoglou2020,farantos2009energy}, but not the bifurcations of the dividing surfaces. The investigation of the bifurcations of dividing surfaces  has been restricted to the case of pitchfork bifurcations  \cite{mauguiere2013bifurcations,katsanikas2021bifurcation,lyu2021hamiltonian} and Morse bifurcations \cite{mackay2014bifurcations} (for more details about the references about the bifurcations of the transition states and dividing surfaces see the introduction of \cite{katsanikas2021bifurcation}).

In this paper, we study for the first time the structure of dividing surfaces before and after a period-doubling bifurcation. Furthermore, we will investigate the range, the minimum, and maximum of the dividing surfaces before and after a period-doubling bifurcation. We will study all the cases of a period-doubling bifurcation, the supercritical and subcritical cases. This is the first time that it is studied the structure of the dividing surfaces that are constructed from periodic orbits with high order multiplicity. 

The description of the potential energy surface is at section \ref{model}. Then we analyze one supercritical and one subcritical period-doubling bifurcation of one of the families of the well (section \ref{sec:per}). We present our results about the corresponding bifurcations of dividing surfaces in section \ref{results} and we discuss our conclusions in section \ref{conclusions}.

\section{Model}
\label{model}

The analytical form of the PES that we study was inspired by \cite{collins2013nonstatistical} and has the form:

\begin{equation}
V(x,y) = c_1(x^2+y^2)+c_2 y-c_3(x^4+y^4-6x^2 y^2)
\label{PES}
\end{equation}

For the parameters $c_{1}=5 , c_{2}=3, c_{3}=-0.3$, the PES has a minimum (well) which is surrounded by 4 index-1 saddles. The energy of the two upper index-1 saddles is higher than the energy  of two lower index-1 saddles. These four index-1 saddles control the entrance and the exit from the central area of the caldera. A three dimensional plot of the PES is given in Fig. \ref{PES}. The two dimensional (2D) Hamiltonian is given by

\begin{equation}\label{hamil}
  H(x,y,p_x,p_y) =\frac{p_x^2}{2m}+\frac{p_y^2}{2m}+V(x,y), 
\end{equation}

\noindent
where we consider $m$ to be a constant equal to $1$. The corresponding Hamiltonian vector field (i.e. Hamilton's equations of motion) is:

\begin{equation}\label{ham1}
    \begin{aligned}
    &\dot{x} = \frac{\partial H}{\partial p_x} = \frac{p_x}{m}\\
    &\dot{y} = \frac{\partial H}{\partial p_y} = \frac{p_y}{m}\\
    &\dot{p_x} = - \frac{\partial V}{\partial x}(x,y) = -(2c_1x-4c_3x^3+12c_3xy^2)\\
    &\dot{p_y} = - \frac{\partial V}{\partial y}(x,y) = -(2c_1y-4c_3y^3+12c_3x^2y+c_2)
    \end{aligned}
\end{equation}

\noindent
Hamilton's equations conserve the total energy, which we will denote as $E$ throughout the paper.

\begin{figure}[htbp] 
	\begin{center}
		\includegraphics[scale=0.5]{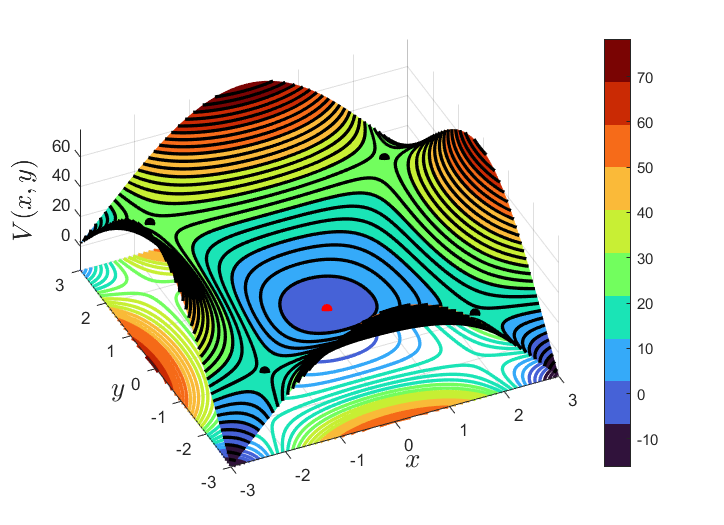} 
	\caption{Plot of the PES given in Eq. (\ref{PES}).}
    \end{center}
    \label{Pess}
\end{figure}

\begin{figure}[htbp] 
	\begin{center}
		\includegraphics[scale=0.5]{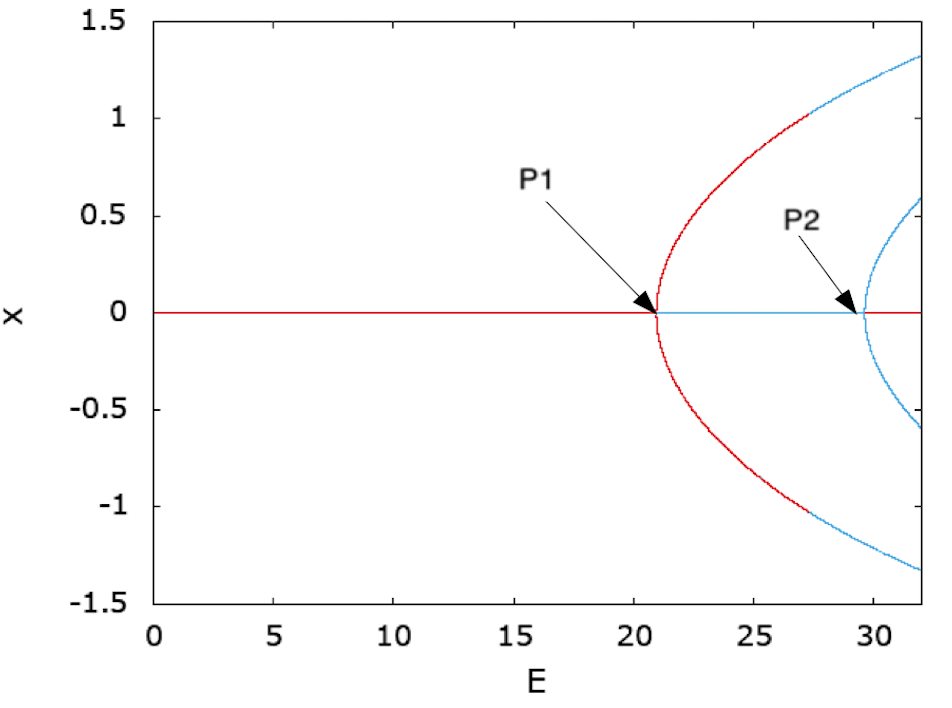} 
	\caption{The coordinate $x$ (in the Poincar\'e section $y=0$ with $p_y>0$) of the periodic orbits of the family of the well and its bifurcations. The stable and unstable parts of the families are depicted by  red and cyan colors, respectively. P1 and P2 are the points of bifurcation for the supercritical and subcritical cases, respectively.}
    \end{center}
    \label{Pess1}
\end{figure}

\section{Periodic Orbits}
\label{sec:per}

\begin{figure}
 \centering
A)\includegraphics[scale=0.33]{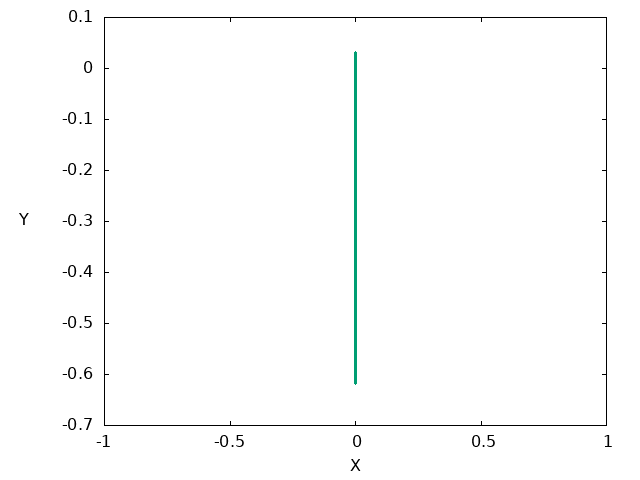}
B)\includegraphics[scale=0.33]{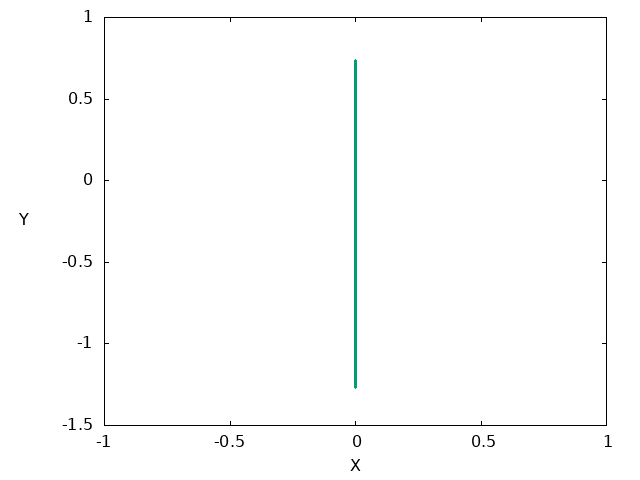}\\
C)\includegraphics[scale=0.33]{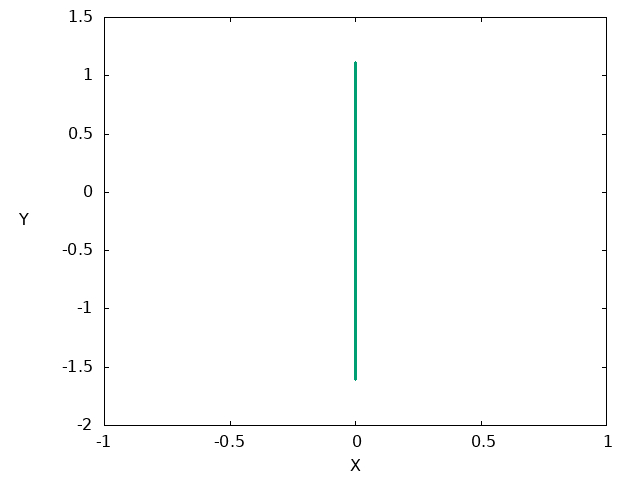}
D)\includegraphics[scale=0.33]{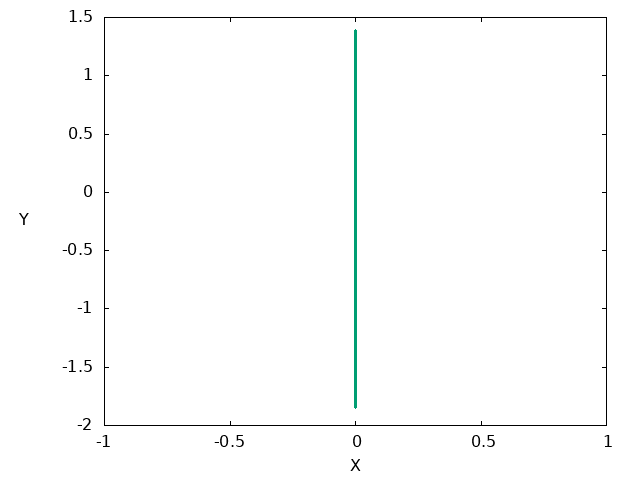}\\
E)\includegraphics[scale=0.33]{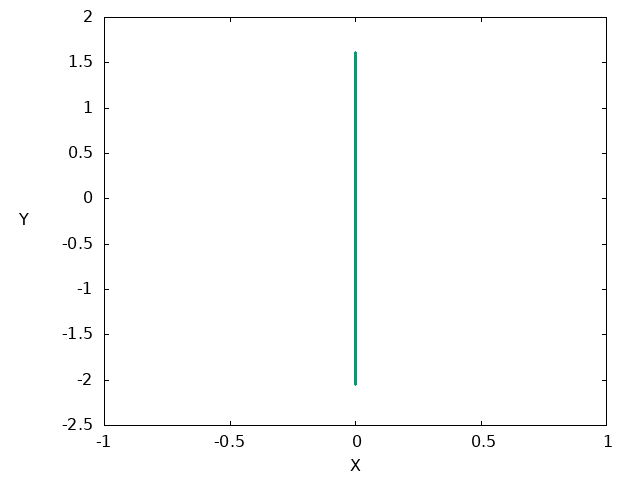}
F)\includegraphics[scale=0.33]{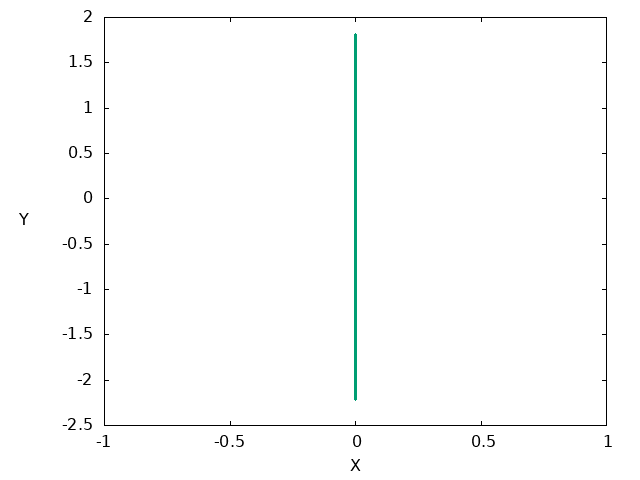}\\
G)\includegraphics[scale=0.33]{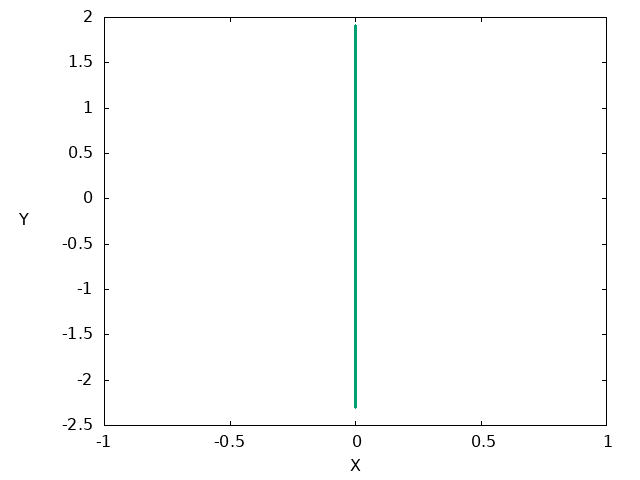}
H)\includegraphics[scale=0.33]{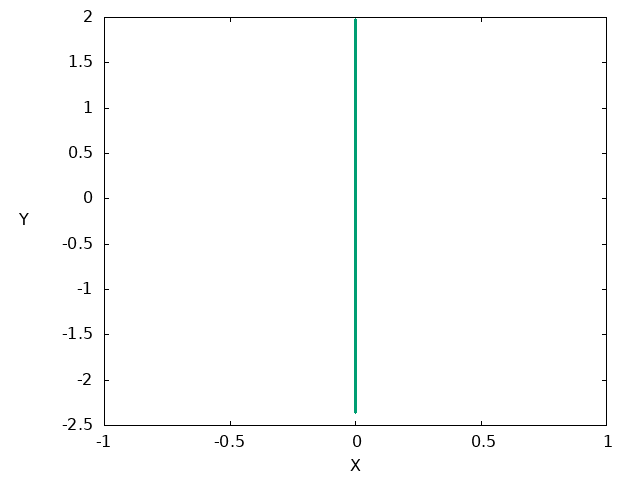}\\
I)\includegraphics[scale=0.33]{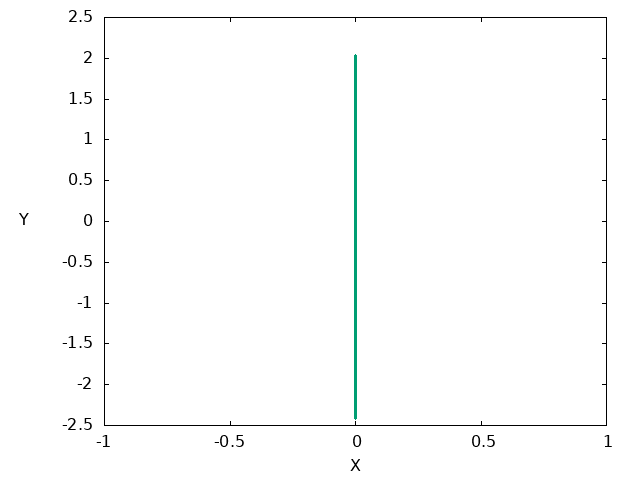}
J)\includegraphics[scale=0.33]{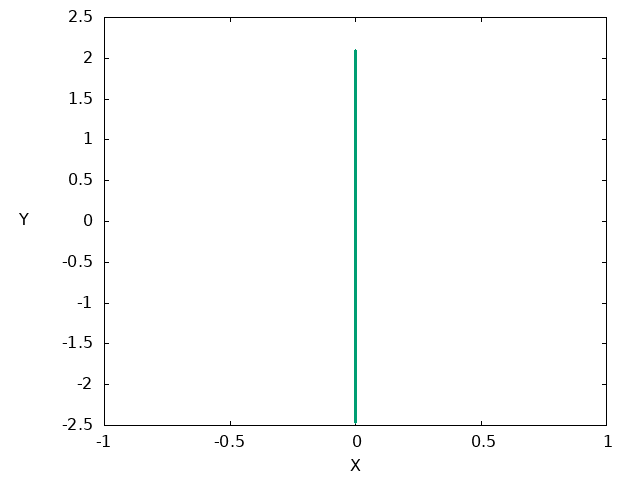}\\
\caption{2D projections of the periodic orbits of the  family of the well in the configuration space for energies A) 0.1 B) 5 C) 10 D) 15 E) 20 F) 25 G) 28 H) 30 I) 32 J) 34.}
\label{periodicorbitscentralxy}
\end{figure}

\begin{figure}
 \centering
A)\includegraphics[scale=0.33]{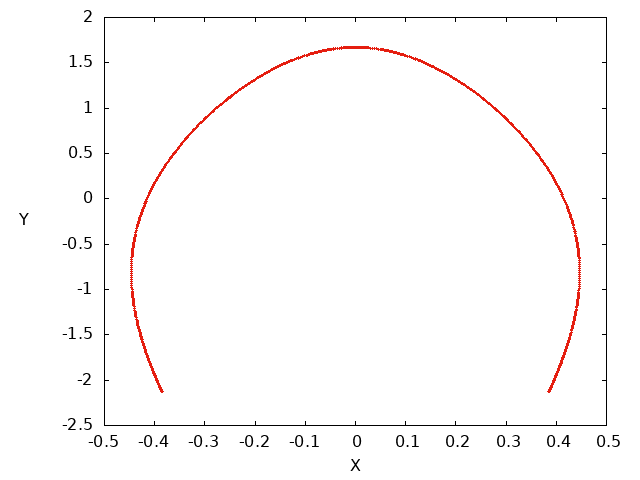}
B)\includegraphics[scale=0.33]{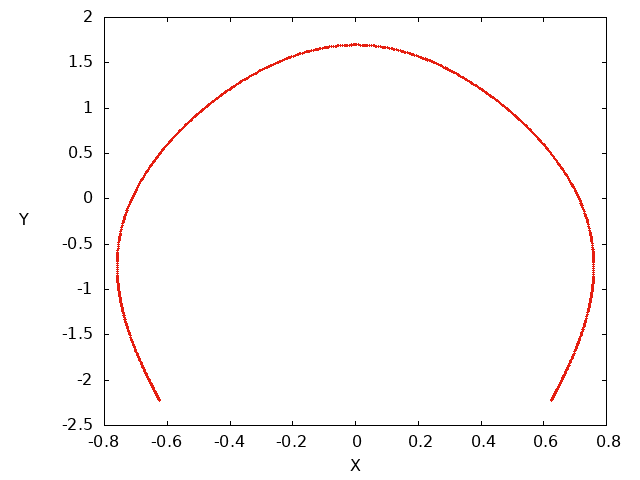}\\
C)\includegraphics[scale=0.33]{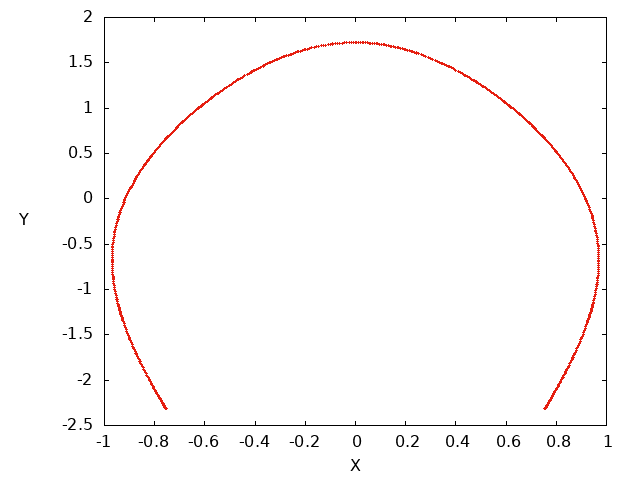}
D)\includegraphics[scale=0.33]{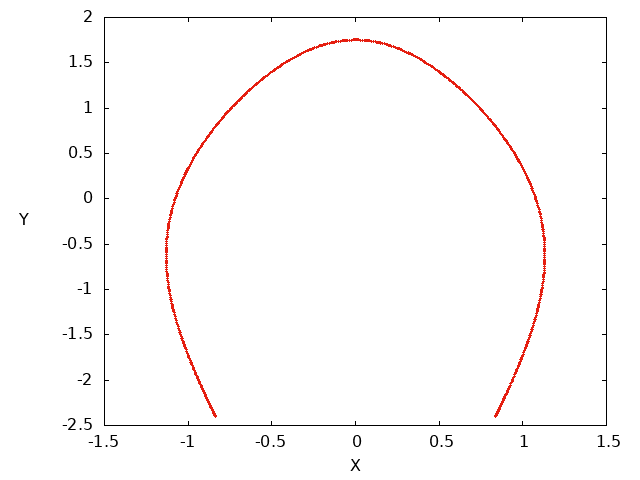}\\
E)\includegraphics[scale=0.33]{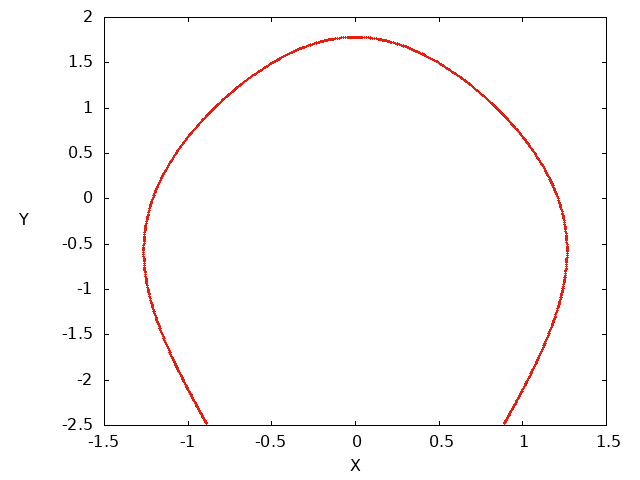}
F)\includegraphics[scale=0.33]{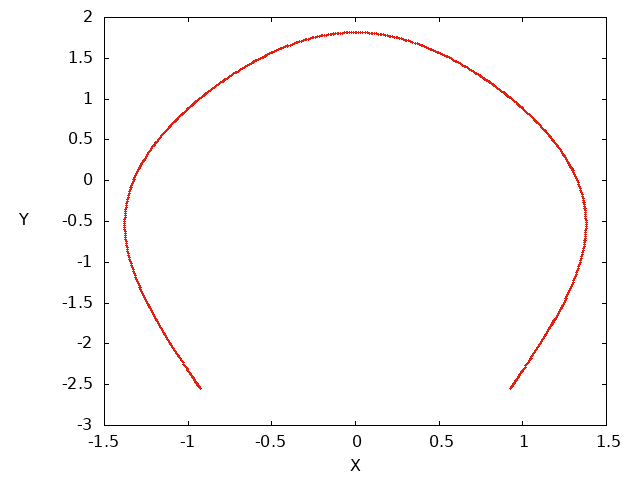}\\
G)\includegraphics[scale=0.33]{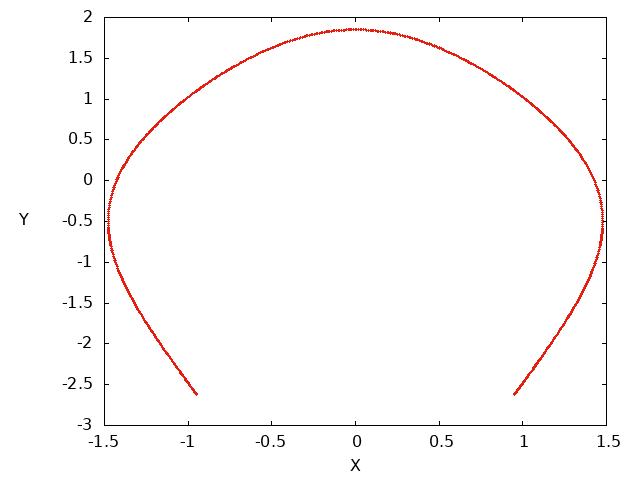}
\caption{2D projections of the periodic orbits of the first period-doubling bifurcation of the  family of the well in the configuration space for energies A) 22 B) 24 C) 26 D) 28 E) 30 F) 32 G) 34.}
\label{periodicorbitsperdoubxy}
\end{figure}

\begin{figure}
 \centering
A)\includegraphics[scale=0.33]{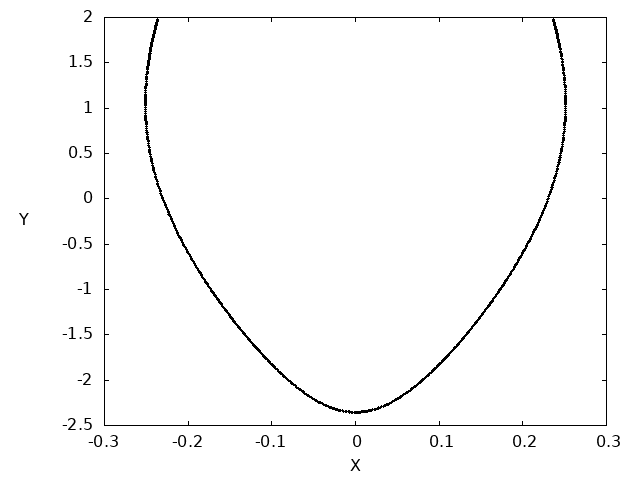}
B)\includegraphics[scale=0.33]{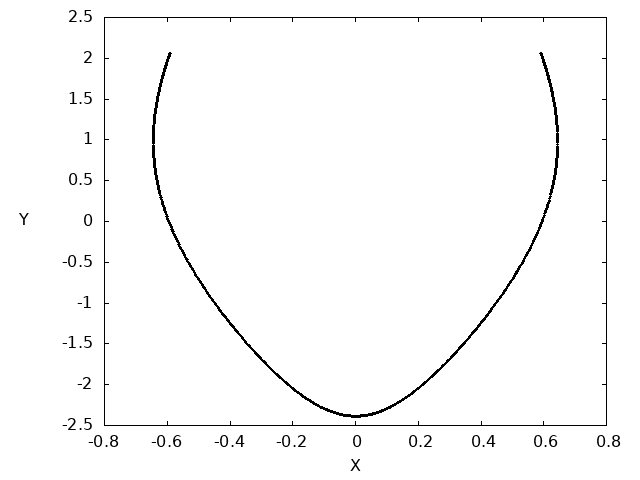}\\
C)\includegraphics[scale=0.33]{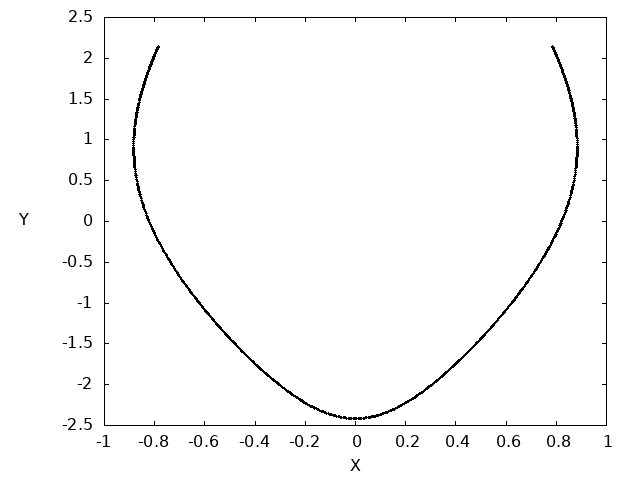}
\caption{2D projections of the periodic orbits of the second period-doubling bifurcation of the  family of the well in the configuration space for energies A) 30 B) 32 C) 34.}
\label{periodicorbitsperdsxy}
\end{figure}

As we described in the introduction, we have a central minimum on the potential energy surface of our model. According to the Lyapunov subcenter theorem 
we have two families of periodic orbits associated with this minimum, or ''well'',  (see \cite{rabinowitz1982periodic,weinstein1973normal,moser1976periodic}). In this section, we describe the period-doubling bifurcations of one of these families of periodic orbits of the well (the family of periodic orbits with period 1 -see \cite{katsanikas2018phase} for more details). This family was initially stable and then it has two period-doubling bifurcations, one supercritical at the point P1 (see Fig. \ref{Pess1} - at this point the basic family becomes unstable) and one subcritical at the point P2 (see Fig.\ref{Pess1} - at this point the basic family becomes stable). The supercritical and subcritical period-doubling bifurcation will be referred to in the paper as the first and second period-doubling bifurcation of the family of the well, respectively, this is because of their order of appearance. The periodic orbits of the family of the well are vertical lines in the configuration space (see Fig. \ref{periodicorbitscentralxy}), The length of these lines is increased as the energy increases (see Fig. \ref{periodicorbitscentralxy}). The periodic orbits of the first and second period-doubling bifurcations have a similar geometry in the configuration space with that of horseshoe curves  (see Figs.\ref{periodicorbitsperdoubxy}, \ref{periodicorbitsperdsxy}). These curves are directed downward and upward in the case of the first and second period-doubling bifurcations, respectively. The extension of these curves in the $x-$ and $y-$ directions increases as the energy increases.

%2D projections of periodic orbits  
%X-Y
%(Main (Central))

%(PERIOD doubling)

%(ds)

\section{Results}
\label{results}

Here we focus our study initially on the evolution of the structure of the dividing surfaces before and after a supercritical and a subcritical period-doubling bifurcation of one family of the well of our model. The algorithm for the construction of the dividing surfaces was given in the appendix (see \ref{appen}).  First we study the structure of the dividing surfaces that correspond to the periodic orbits of the family of the well before the first period-doubling bifurcation (supercritical - before the point P1 in Fig. \ref{Pess1}). Second, we study the structure of the dividing surfaces associated with the periodic orbits of the first period-doubling bifurcation and their position with respect to the position of the dividing surfaces associated with the periodic orbits of the family of the well (after the point P1 and before the point P2 in Fig. \ref{Pess1}). Then, we study the structure of dividing surfaces of the second period-doubling bifurcation(subcritical - after the point P2 in Fig. \ref{Pess1})  and their position with respect to the dividing surfaces with respect to the position of the dividing surfaces associated with the periodic orbits of the family of the well and the first period-doubling bifurcation. Moreover, we study the range, the minimum, and maximum of the dividing surfaces as the energy varies in all cases.

\begin{figure}
 \centering
A)\includegraphics[scale=0.33]{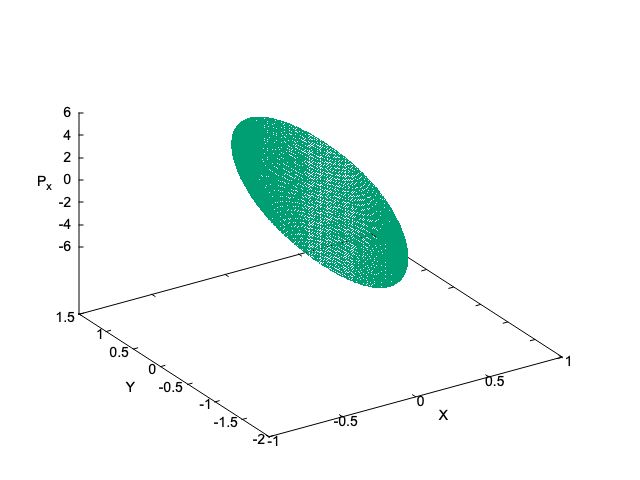}
B)\includegraphics[scale=0.33]{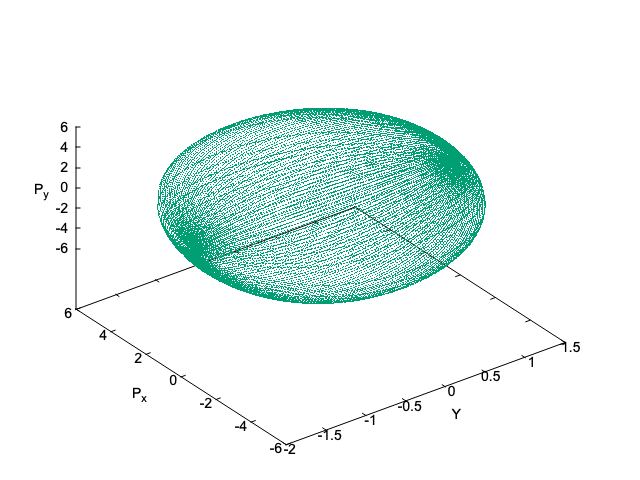}\\
C)\includegraphics[scale=0.33]{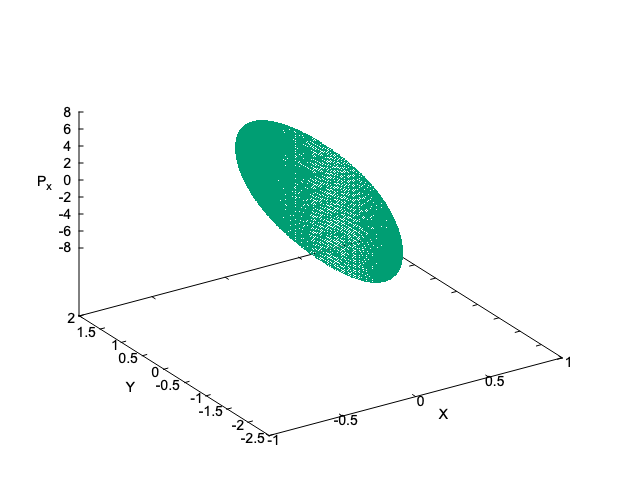}
D)\includegraphics[scale=0.33]{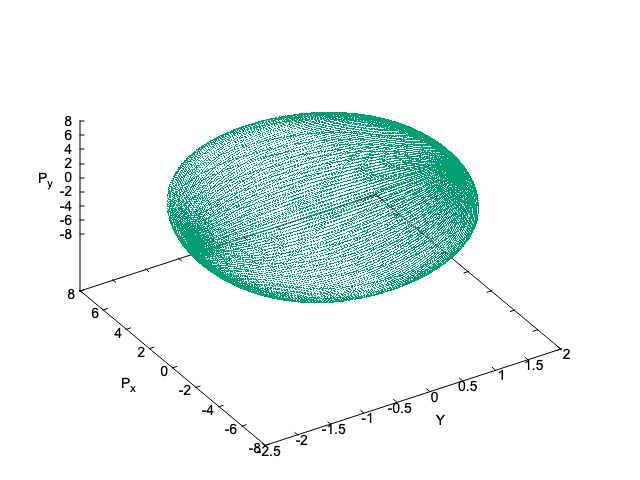}\\
E)\includegraphics[scale=0.33]{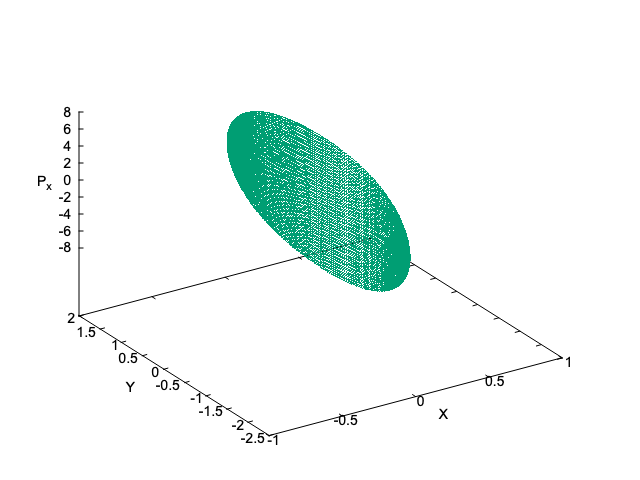}
F)\includegraphics[scale=0.33]{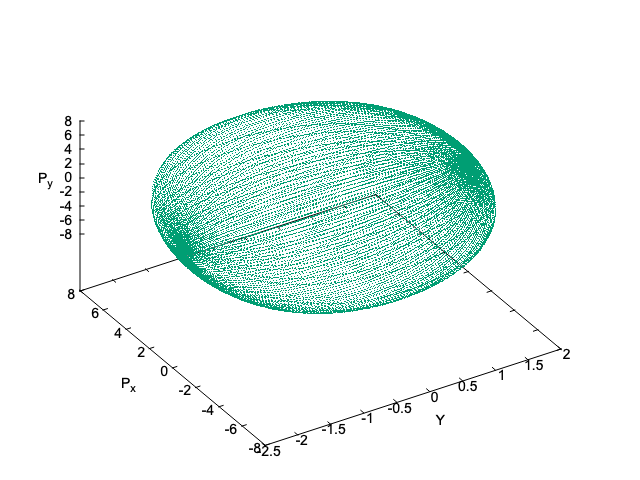}\\
\caption{The 3D projections $(x,y,p_x)$   (panels A, C and E for values of energy 15, 24 and 30 respectively - first column) and  and $(y,p_x,p_y)$ (panels B, D and F for values of energy 15, 24 and 30 respectively - second column) of the dividing surfaces associated with the periodic orbits of the family of the well. } 
\label{div-central}
\end{figure}

First, we study the structure of dividing surfaces of the periodic orbits of the well before the two period-doubling bifurcations. We see that the dividing surfaces (for value of energy 15) are presented as ellipsoids in the 3D projection $(y,p_x,p_y)$ (see panel B of Fig. \ref{div-central}) and as filamentary structures in the 3D projections $(x,y,p_x)$, $(x,y,p_y)$ and $(x,p_x,p_y)$ of the phase space (see for example panel A of Fig. \ref{div-central}). Then as we increase the energy we see that  the filamentary structures extend more and more in the $y$-, $p_x$- and $p_y$- directions as we increase the energy (see panels C and E of Fig. \ref{div-central} for values of energy above the two period-doubling bifurcations). We observe the same extension in these directions for the ellipsoids in the 3D subspace $(y,p_x,p_y)$ of the phase space (see the panels D and F of Fig. \ref{div-central}). We note that the dividing surfaces of the period-1 family in the Hamiltonian model that we studied in \cite{katsanikas2021bifurcation} were presented as ellipsoids in the 3D subspace $(x,p_x,p_y)$ (and not in the 3D subspace $(y,p_x,p_y)$) and as filamentary structures in the other 3D subspaces of the phase space. This happens because the periodic orbits of the family of the well (that we study in this paper) lie on the $y$-axis in the configuration space and the periodic orbits of the basic family (family of the lower saddles) of the system, that we studied in \cite{katsanikas2021bifurcation}, lie on the $x$- axis in the configuration space. 

\begin{figure}
 \centering
A)\includegraphics[scale=0.33]{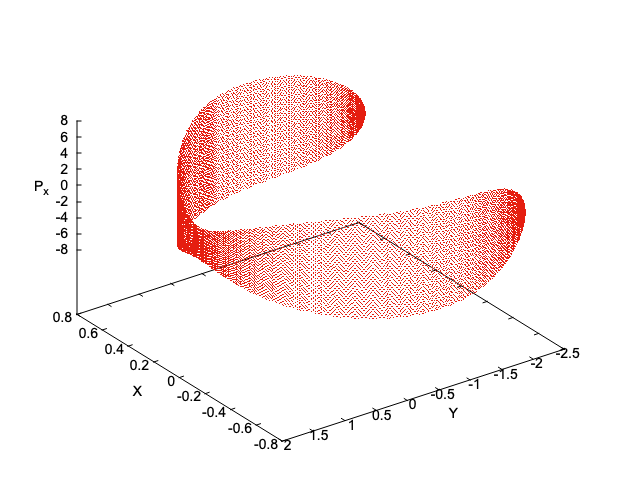}
B)\includegraphics[scale=0.33]{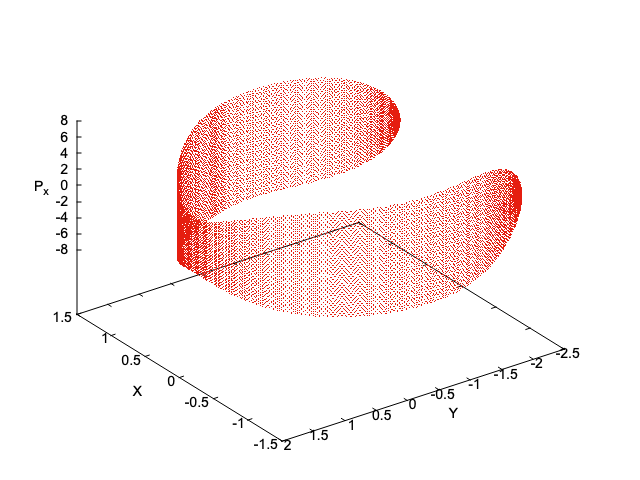}\\
C)\includegraphics[scale=0.33]{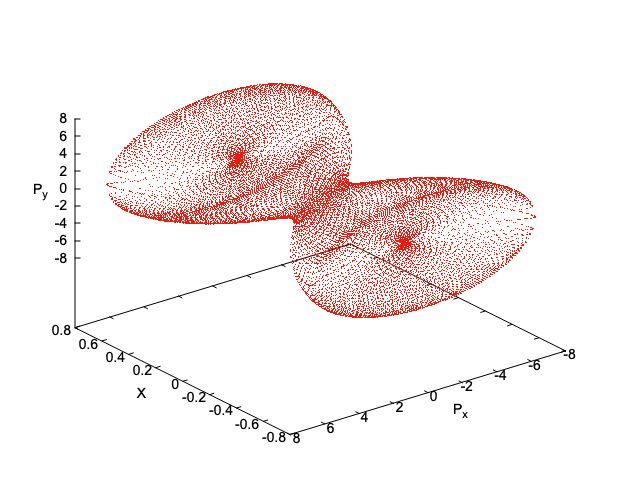}
D)\includegraphics[scale=0.33]{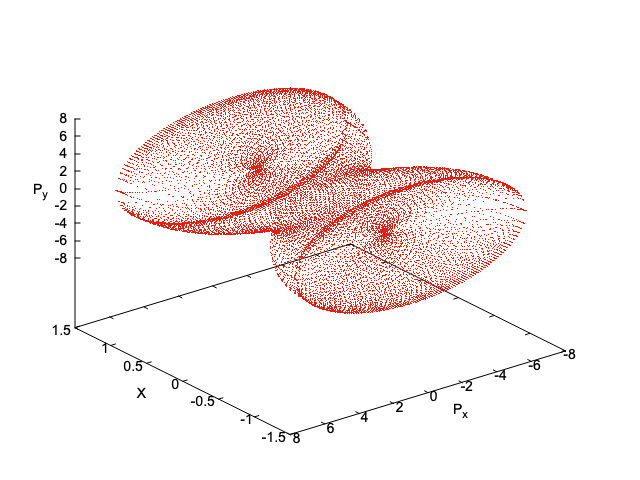}\\
E)\includegraphics[scale=0.33]{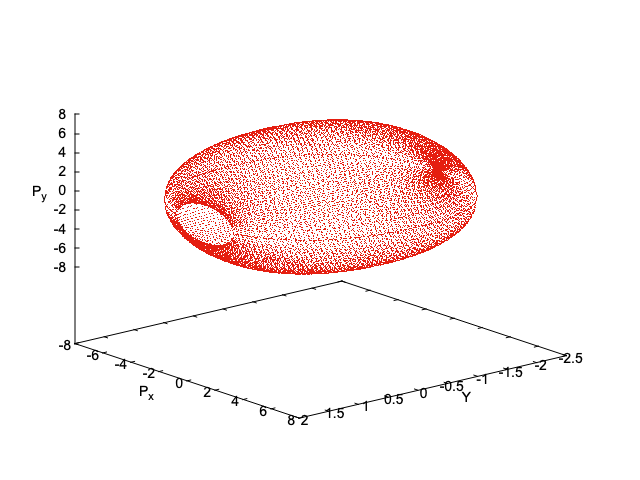}
F)\includegraphics[scale=0.33]{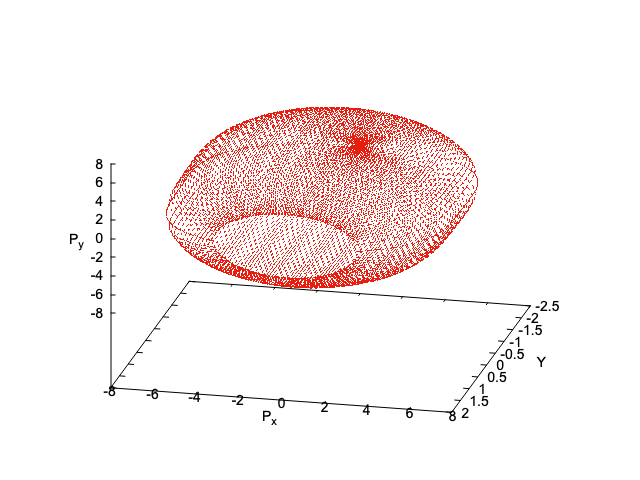}\\
\caption{The 3D projections $(x,y,p_x)$   (panels A and B - first row), $(x,p_x,p_y)$ (panels C and D - second row) and $(y,p_x,p_y)$ (panels E and F - third row) of the dividing surfaces associated with the periodic orbits of the first period-doubling bifurcation. The first column (panels A, C and E) and the second column (panels B, D and F) are for values of energy 24 and 30 respectively. } 
\label{div-bifa}
\end{figure}

Then, we study the structure of the dividing surfaces of the periodic orbits of the first period-doubling bifurcation of the family of the well. These dividing surfaces are presented as  parabolic cylinders in 3D projections $(x,y,p_x)$ and $(x,y,p_y)$   (see for example the  panels A and B of Fig.\ref{div-bifa}). These parabolic cylinders have two branches that move towards the negative $y$ semi-axis. As we increase the energy we see that these branches become more open (they increase the distance between each other in the $x$-direction). Furthermore, the dividing surfaces have similar morphology with this of cylinders without holes in the 3D projection $(x,p_x,p_y)$ of the phase space (see panels C and D of Fig. \ref{div-bifa}).  These cylinders are distorted at the central area close to the plane $x=0$. This distortion becomes smaller as we increase the energy (compare the panels C and D  of Fig. \ref{div-bifa}). The length of these cylinders in the $x$-direction becomes larger as the energy increases (compare the panels C and D  of Fig. \ref{div-bifa}). The dividing surfaces of the periodic orbits of the first period-doubling bifurcation are represented as ellipsoidal rings in the 3D subspace $(y,p_x,p_y)$ of the phase space (see panels E and F of Fig. \ref{div-bifa}). These ellipsoidal  rings are characterized as ellipsoids with a hole (that is found in the positive $y$ semi-axis) that becomes larger as the energy increases (compare the panels E and F  of Fig. \ref{div-bifa}).

\begin{figure}
 \centering
A)\includegraphics[scale=0.33]{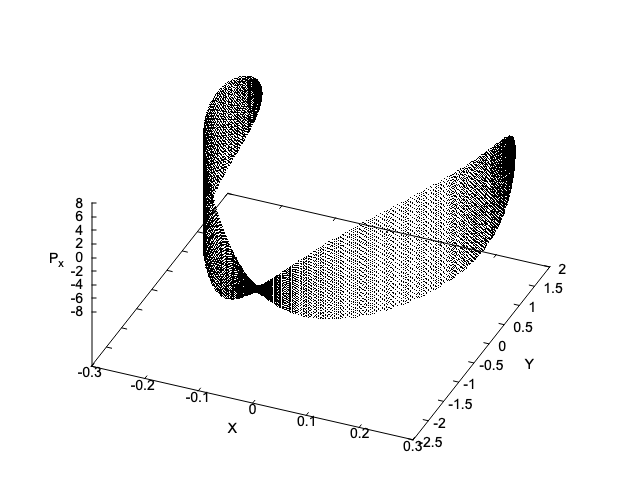}
B)\includegraphics[scale=0.33]{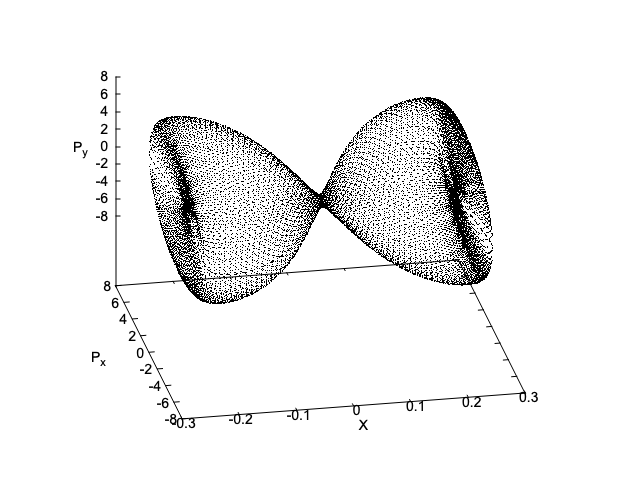}\\
C)\includegraphics[scale=0.33]{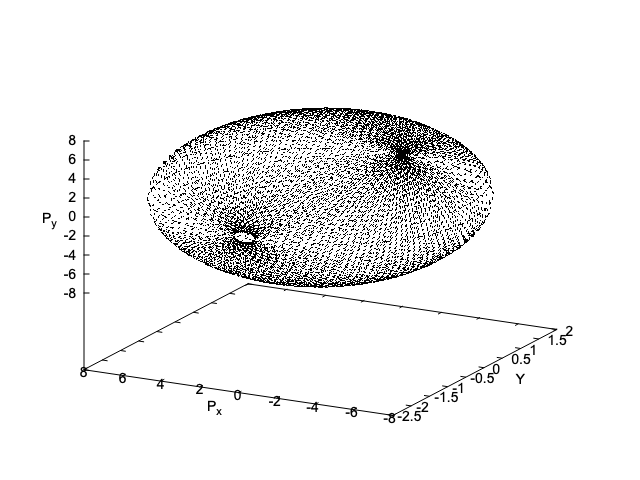}
\caption{The 3D projections $(x,y,p_x)$   (panel A), $(x,p_x,p_y)$ (panel B) and $(y,p_x,p_y)$ (panel C) of dividing surfaces associated of the second period-doubling bifurcation for value of energy 30.} 
\label{div-bifb}
\end{figure}

The next step was to study the structure of dividing surfaces of the periodic orbits of the second period-doubling bifurcation. These objects (see Fig.\ref{div-bifb}) have  the same topology  as the dividing surfaces of the periodic orbits of the first period-doubling bifurcation with two basic  differences:

\begin{enumerate}

\item The branches of the parabolic cylinders have different directions from those of the parabolic cylinders that we described at the first period-doubling bifurcation. They are extended in the positive $y$ semi-axis (see panel A of Fig. \ref{div-bifb}).

\item The ellipsoidal rings of the second period-doubling bifurcation have the hole in the opposite direction in the $y$ axis (it is found for negative $y$ values) to the direction of the ellipsoidal rings of the first period-doubling bifurcation (see the panel C of Fig. \ref{div-bifb} and compare it with the panels E and F  of Fig. \ref{div-bifa}).
\end{enumerate}

\begin{figure}
 \centering
A)\includegraphics[scale=0.33]{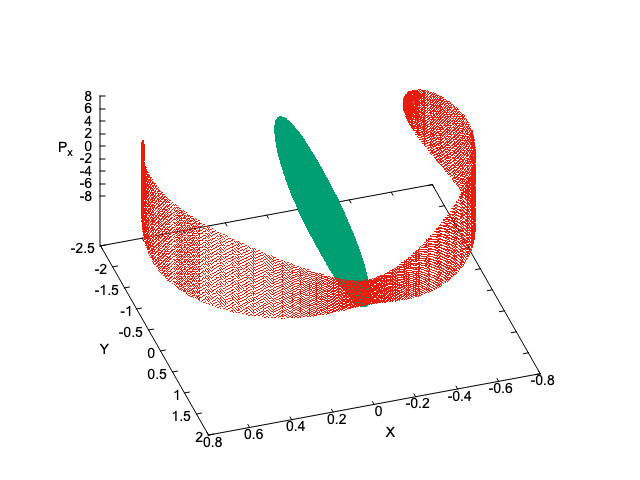}
B)\includegraphics[scale=0.33]{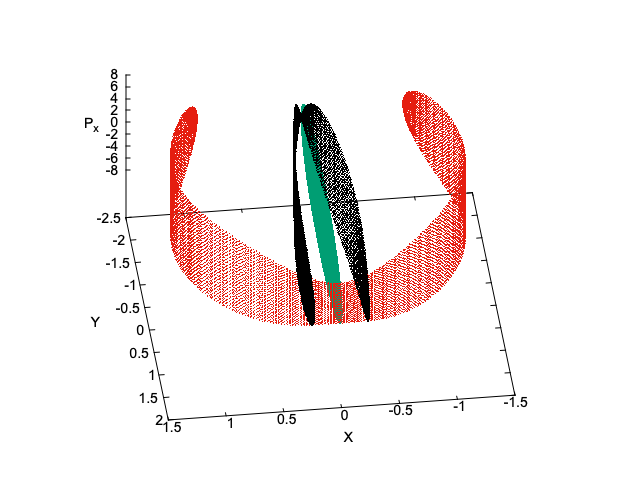}\\
C)\includegraphics[scale=0.33]{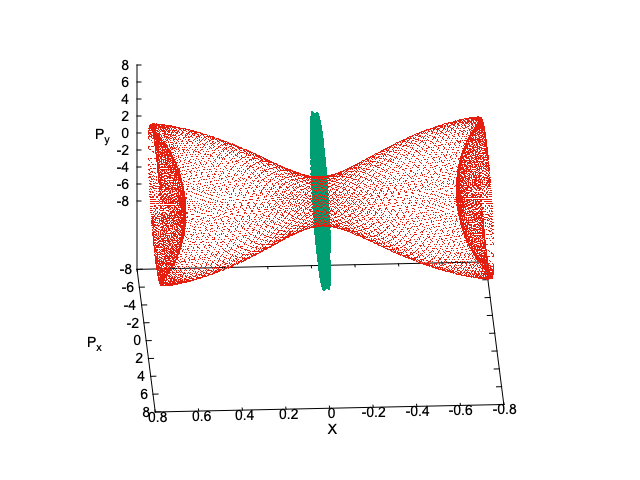}
D)\includegraphics[scale=0.33]{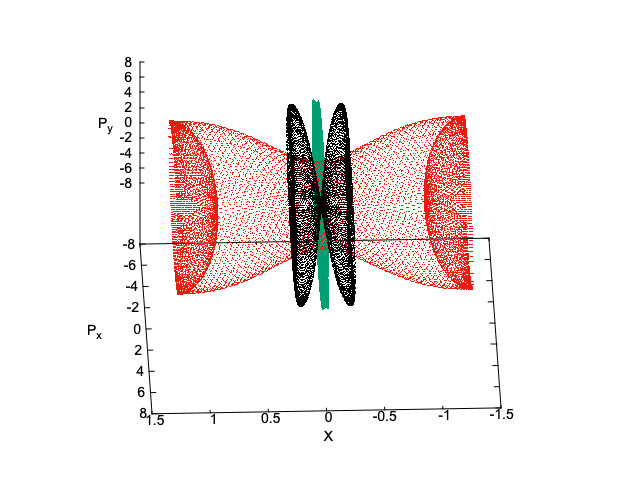}
\caption{The relative positions of the dividing surfaces associated with the periodic orbits of the well (with green color) and the periodic orbits of the first period-doubling bifurcation of the family of the well (with red color) after the first period-doubling bifurcation (for value of energy $E=24$). The relative positions of the dividing surfaces associated with the periodic orbits of the well (with green color), the periodic orbits of the first period-doubling bifurcation of the family of the well (with red color) and the periodic orbits of the second period-doubling bifurcation of the family of the well (with black color) after the second period-doubling bifurcation (for value of energy $E=30$). } 
\label{relative}
\end{figure}

We investigated the relative positions of the dividing surfaces after the two period-doubling bifurcations. Firstly, we analyzed the relative positions of the dividing surfaces in the 3D projection $(x,y,p_x)$ after the first and second period-doubling bifurcations. After the first period-doubling bifurcation, we see the branches of the parabolic cylinder of the periodic orbits of the first period-doubling bifurcation to be extended on both sides of the filamentary structure of the family of periodic orbits of the family of the well (see panel A of Fig. \ref{relative}). After the second period-doubling bifurcation, except the parabolic cylinder of the first period-doubling bifurcation and the filamentary structure of the family of the well, we see another parabolic cylinder, associated with the periodic orbits of the second period-doubling bifurcation, whose branches are extended on both sides of the filamentary structure (see panel B of Fig. \ref{relative}).
These branches lie in the opposite direction to that of the branches of the parabolic cylinder of the first period-doubling bifurcation. The same situation occurs in the 3D projection $(x,y,p_y)$ because the dividing surfaces of the family of the well and of the bifurcating families have similar morphology with this in the  3D projection $(x,y,p_x)$.

Furthermore, we analyzed the relative positions of the dividing surfaces in the 3D subspace $(x,p_x,p_y)$ of the phase space. In this 3D subspace, we see the
filamentary structure (associated with the family of the well), that lies in the plane $x=0$, to be at the central area of the  
distorted cylinder without holes, associated with the periodic orbits of the first period-doubling bifurcation (see panel C of Fig. \ref{relative}).  After the second period-doubling bifurcation, we see a second distorted cylinder without holes (associated with the periodic orbits of the second period-doubling bifurcation - see panel D Fig. \ref{relative}). We see that the filamentary structure (associated with the family of the well) is at the central area of both distorted cylinders (without holes). This can explain why the cylinders, associated with the bifurcating families, are distorted at their central area in the plane $x=0$. The dividing surfaces associated with the family of the well and the dividing surfaces of the bifurcating families are represented in the 3D subspace $(y,p_x,p_y)$ by an ellipsoid and ellipsoidal rings, respectively (as we explained above). These objects occupy the same volume in this 3D subspace of the phase space.

Finally, we also studied  other features of dividing surfaces like their minimum, maximum and the range:

\begin{figure}
 \centering
A)\includegraphics[scale=0.33]{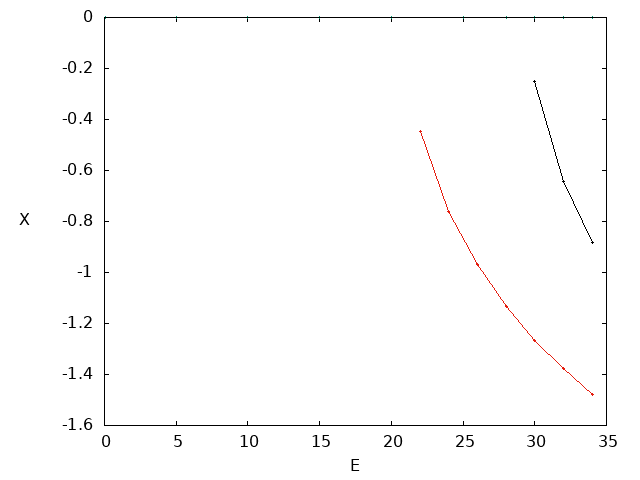}
B)\includegraphics[scale=0.33]{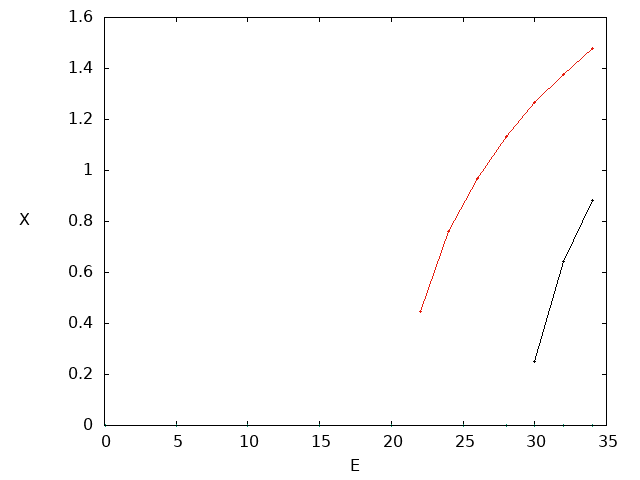}\\
C)\includegraphics[scale=0.33]{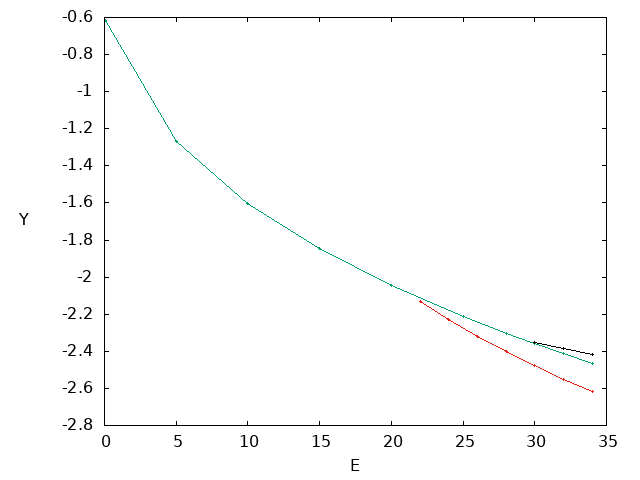}
D)\includegraphics[scale=0.33]{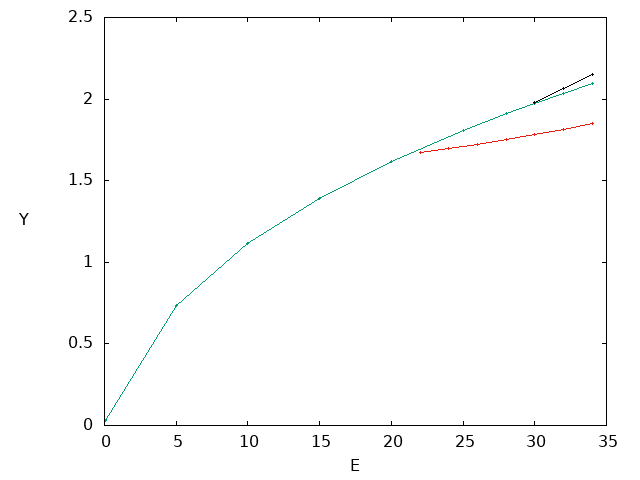}\\
E)\includegraphics[scale=0.33]{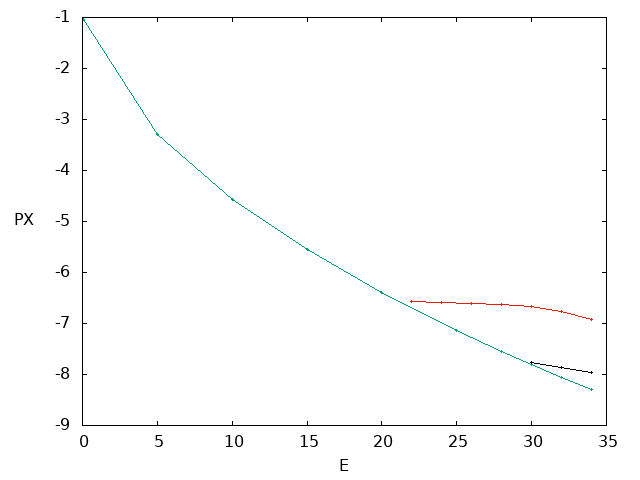}
F)\includegraphics[scale=0.33]{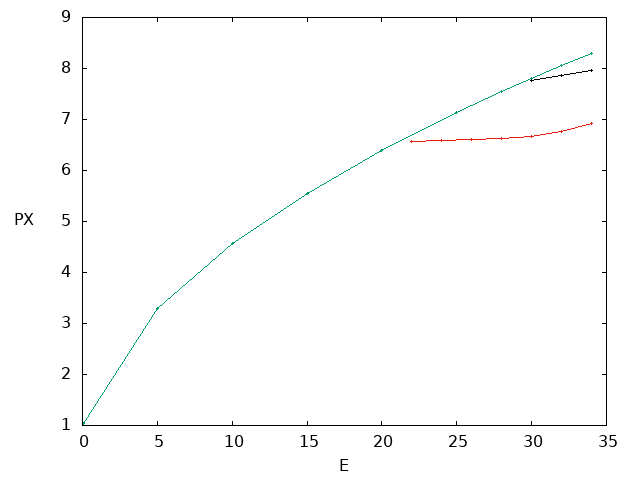}\\
\caption{Diagrams for the A) minimum and B) maximum of the  X coordinate of the family of the well  (green), first  period doubling bifurcation (red) and second period doubling  bifurcation (black) families versus the energy, C) minimum and D) maximum of the  Y coordinate of all families versus the energy and E) minimum and F) maximum of the  PX coordinate of all families versus the energy.}
\label{minmax}
\end{figure}

\begin{figure}
 \centering
A)\includegraphics[scale=0.5]{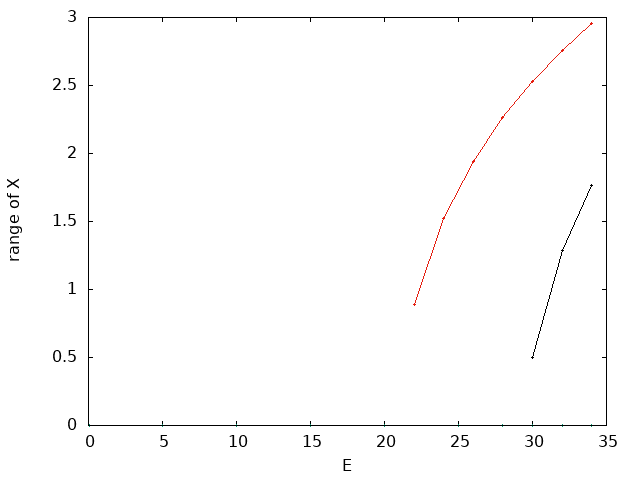}\\
B)\includegraphics[scale=0.5]{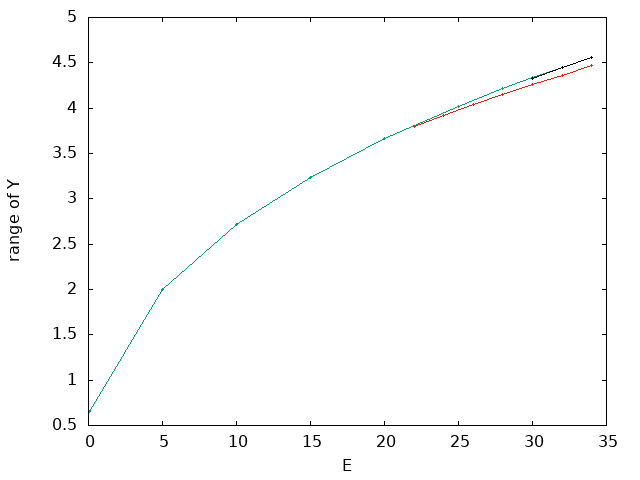}\\
C)\includegraphics[scale=0.5]{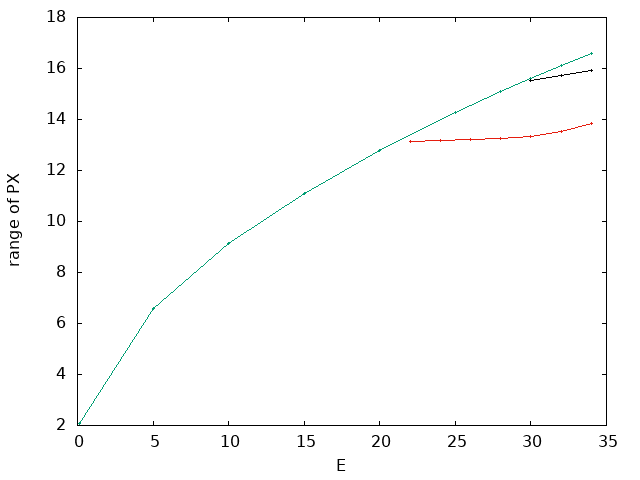}
\caption{A) Range of the X coordinate of the  family of the well (green line), the first period doubling family (red line) and second period-doubling bifurcation family (black line) versus the energy, B) Range of the Y coordinate of all families versus the energy and C) Range of the PX coordinate of all families, versus the energy.}
\label{range}
\end{figure}

\begin{itemize}

\item {\bf Maximum and Minimum of the dividing surfaces:} In Figure \ref{minmax} we visualize the minimum and maximum of the dividing surfaces with respect to the X coordinate in the panels A and B respectively, to the Y coordinate in the panels C and D and finally the minimum and maximum to the PX coordinate in the panels E and F. Notice that the graphs for the minimum and maximum with respect to the PY coordinate are missing since they can be determined from the other three coordinates (X,Y,PX) of the Hamiltonian of the system. All families are depicted in green (the family of the well), in red (the first period-doubling bifurcation) and in black ( the second), versus the energy. In the panels A, C and E we notice that the minimum of the X, Y and PX coordinates respectively of all families is decreasing as we increase the energy. Analogously, in panels B, D and F the maximum of the X, Y and PX coordinates respectively of all families is increasing as we increase the energy. Finally it is important to observe that the minimum and maximum with respect to the X coordinate of the green family (central family) is zero for all energies. In all other cases the two bifurcating families are represented as branches of the initial curve that corresponds to the family of the well.

\item {\bf The range of the dividing surfaces:} In Figure \ref{range} we present the range of the dividing surfaces with respect to the X, Y and PX coordinate respectively of all families in green color (family of the well), red color (first period-doubling bifurcation) and black color (second period-doubling bifurcation). We notice that the range with respect to all coordinates of all families is increasing as we increase the energy. The only exception is in the panel A where the range of the central family (green family) of the X coordinate is zero for all energies. In all other cases the two bifurcating families are represented as branches of the initial curve that corresponds to the family of the well.

\end{itemize}

\begin{figure}
 \centering
A)\includegraphics[scale=0.33]{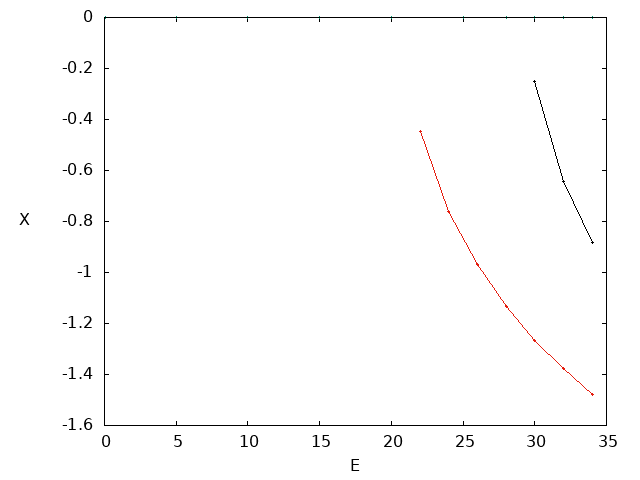}
B)\includegraphics[scale=0.33]{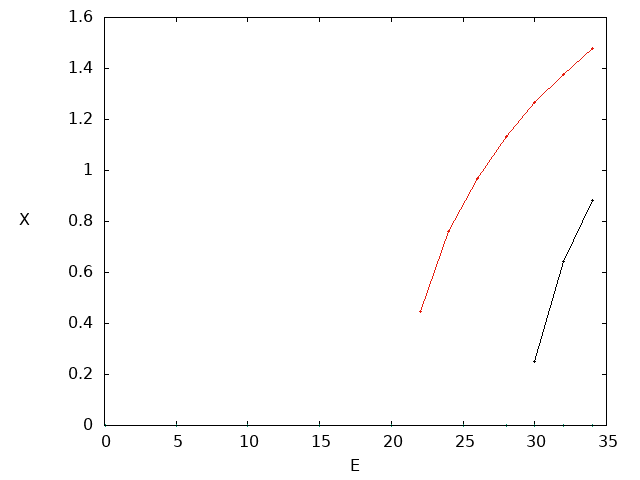}\\
C)\includegraphics[scale=0.33]{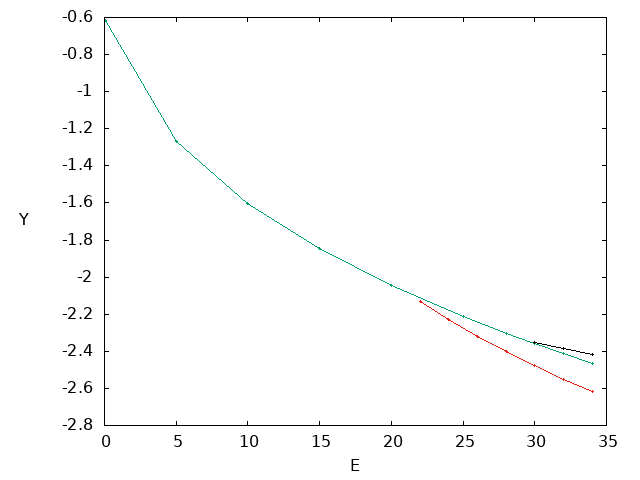}
D)\includegraphics[scale=0.33]{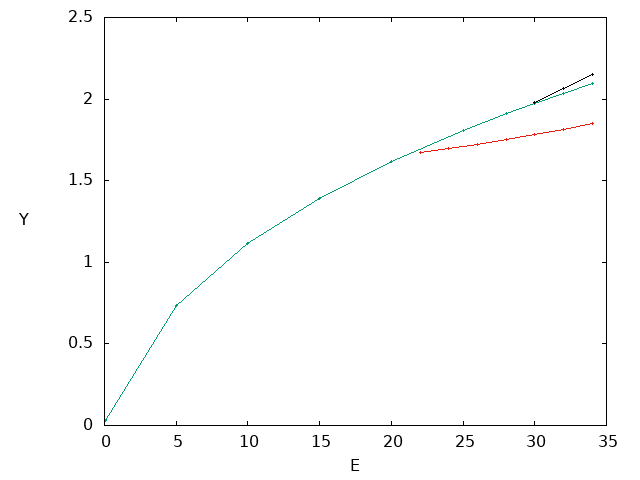}\\
E)\includegraphics[scale=0.33]{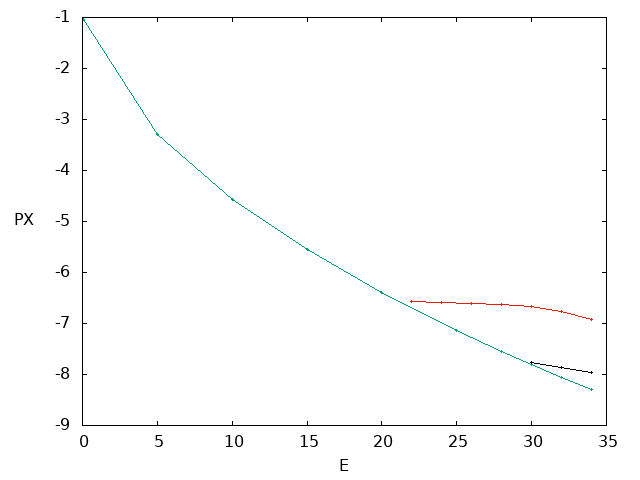}
F)\includegraphics[scale=0.33]{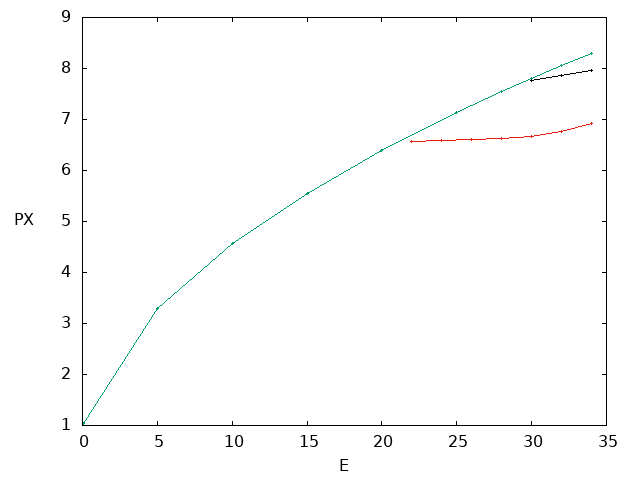}\\
\caption{Diagrams for the A) minimum and B) maximum of the  X coordinate of the family of the well  (green), first  period doubling bifurcation (red) and second period doubling  bifurcation (black) families versus the energy, C) minimum and D) maximum of the  Y coordinate of all families versus the energy and E) minimum and F) maximum of the  PX coordinate of all families versus the energy.}
\label{minmax}
\end{figure}

\section{Conclusions}
\label{conclusions}

We study the geometrical structure of the dividing surfaces, i.e., their  minimum, maximum and range before and after the supercritical and subcritical period-doubling bifurcation of the family of the well of our model (a caldera-type potential).
The structure and the relative position of the dividing surfaces associated with the period-doubling bifurcations in the 3D subspaces of the phase space are:

\begin{enumerate}
    \item Parabolic cylinders that have two branches to be extended on both sides of the dividing surfaces associated with the periodic orbits of the initial family. The difference between the two period-doubling bifurcations for the first (supercritical) and the second (subcritical) is that the branches of the parabolic cylinders are in opposite directions.
    
    \item Ellipsoidal rings that have a hole. The difference between the two period-doubling bifurcations, the first (supercritical) and the second (subcritical), is that the hole of the ellipsoidal rings is on opposite sides of the ellipsoidal ring. 
    
    \item Distorted cylinders without holes. The dividing surfaces of the initial family are at the central region of these cylinders. In this region, we have a distortion of the surface of these cylinders. 
\end{enumerate}

The range, minimum, and maximum of the dividing surfaces of bifurcating families versus energy are represented as branches of the range, minimum and maximum of the dividing surfaces associated with the initial family. 

\section{appendix}
\label{appen}

The construction of periodic orbit dividing surfaces for  Hamiltonian systems with two degrees of freedom was introduced in \cite{Pechukas73,Pechukas77,Pollak78,Pechukas79}. The algorithm for this construction was described in \cite{waalkens2004direct, ezra2018sampling,haigh2021time}. Here we give a brief outline of the algorithm for completeness.

\begin{enumerate} 

\item Locate a periodic orbit.

\item Project the periodic orbit into configuration space.

\item Choose points on that curve $(x_i,y_i)$ for $i=1,...N$ where  $N$ is the desired number of points.  Points are spaced uniformly according to distance along the periodic orbit. 

\item For each point  $(x_i,y_i)$ determine $p_{x max,i}$ by solving:

\begin{eqnarray}
\label{eq3a}
 H(x_i, y_i, p_x, 0)=\frac {p_x^2}{2m}+ V(x_i, y_i)
\end{eqnarray}

\noindent
for $p_x$. Note that a solution of this equation requires $E-V(x_i,y_i) \geq 0$ and that there will be two solutions, $\pm p_{x max,i}$. 
\item For each point $(x_i,y_i)$ choose points for $j=1,...,K$ with $p_{x_1}=-p_{x max,i}$ and  $p_{x_K}=p_{x max,i}$  and solve the equation  $H(x_i, y_i, p_x, p_y)=E$ to obtain $p_y$ (we will obtain two solutions  $p_y$, one negative and one positive).

\end{enumerate}

This algorithm gives produces a phase space dividing surface with the desired properties. It satisfies the no-recrossing property and the phase space flux across the dividing surface is a minimum with respect to other possible dividing surfaces.

\section*{Acknowledgments}
The authors acknowledge the financial support provided by the EPSRC Grant No. EP/P021123/1 and MA acknowledges support from the grant CEX2019-000904-S and IJC2019-040168-I funded by: MCIN/AEI/ 10.13039/501100011033.

\clearpage
\bibliography{SNreac}

\end{document}